\documentclass[a4paper,reprint,nofootinbib,superscriptaddress,onecolumn,notitlepage]{revtex4-1}

\usepackage[utf8]{inputenc}
\usepackage[T1]{fontenc}

\usepackage{graphicx}
\usepackage{amsmath}
\usepackage{mathrsfs}  
\usepackage{braket}
\usepackage{amsfonts} 
\usepackage{amssymb} 
\usepackage{array}
\usepackage{booktabs}
\usepackage{multirow}
\usepackage{csquotes}
\usepackage{mathtools}
\usepackage[dvipsnames]{xcolor}

\usepackage[toc,title]{appendix}
\usepackage{siunitx}

\definecolor{ct_black}{HTML}{000000}
\definecolor{ct_orange}{HTML}{ED872D}
\definecolor{ct_purple}{HTML}{7A68A6}
\definecolor{ct_blue}{HTML}{348ABD}
\definecolor{ct_turquoise}{HTML}{188487}
\definecolor{ct_red}{HTML}{E32636}
\definecolor{ct_pink}{HTML}{CF4457}
\definecolor{ct_green}{HTML}{467821}

\definecolor{tab_blue}{HTML}{1F77B4}
\definecolor{tab_orange}{HTML}{FF7F0E}
\definecolor{tab_green}{HTML}{2CA02C}
\definecolor{tab_red}{HTML}{D62728}
\definecolor{tab_purple}{HTML}{9467BD}
\definecolor{tab_brown}{HTML}{8C564B}
\definecolor{tab_pink}{HTML}{E377C2}
\definecolor{tab_gray}{HTML}{7F7F7F}
\definecolor{tab_olive}{HTML}{BCBD22}
\definecolor{tab_cyan}{HTML}{17BECF}

\usepackage{hyperref}
\usepackage{bookmark}

\hypersetup{%
  breaklinks=true,
  colorlinks=true,
  linkcolor=ct_green,
  filecolor=ct_green,
  urlcolor=ct_green,
  citecolor=ct_green,
  pdfborder=0 0 0,
}

\def\det{\ensuremath{\operatorname{det}}}

\newcommand{\ee}{{\rm e}}
\newcommand{\ii}{{\rm i}}

\def\Id{\ensuremath{\text{Id}}}

\newcommand{\strong}[1]{\textbf{#1}}

\RequirePackage{expl3,xparse,l3regex}
\ExplSyntaxOn
\NewDocumentCommand{\hm}{m}
{
    \tl_set:Nx \l_hm_contents {\ensuremath{#1}}
    \regex_replace_all:nnN { -(.) } { \c{overline}\cB\{ \1 \cE\} } \l_hm_contents
    \regex_replace_all:nnN { (.)\_(.) } { \1\c{c_math_subscript_token}\2 } \l_hm_contents
    \regex_replace_all:nnN { ([a-zA-Z]+) } { \c{text}\cB\{ \1 \cE\} } \l_hm_contents
    \tl_use:N \l_hm_contents
}

\NewDocumentCommand{\sch}{m}
{
    \tl_set:Nx \l_sch_contents {#1}
    \regex_replace_all:nnN { \_ }{} \l_sch_contents 
    \regex_replace_all:nnN { (.)(.*) } { \1\c{c_math_subscript_token}\cB\{\2\cE\} } \l_sch_contents 
    \regex_replace_all:nnN { (d|h|v|s) }{ \c{text}\cB\{\1\cE\} }  \l_sch_contents 
    \regex_replace_all:nnN { \cS\  }{} \l_sch_contents 
    \tl_set:Nn \l_sch_result {\ensuremath{\l_sch_contents}} 
    \tl_use:N \l_sch_result
}

\NewDocumentCommand{\irrep}{m}
{
    \tl_set:Nx \l_sch_contents {#1}
    \regex_replace_all:nnN { \_ }{} \l_sch_contents 
    \regex_replace_all:nnN { (.)(.*) } { \1\c{c_math_subscript_token}\cB\{\2\cE\} } \l_sch_contents 
    \regex_replace_all:nnN { ([a-zA-Z]) }{ \c{text}\cB\{\1\cE\} }  \l_sch_contents 
    \regex_replace_all:nnN { \cS\  }{} \l_sch_contents 
    \tl_set:Nn \l_sch_result {\ensuremath{\l_sch_contents}} 
    \tl_use:N \l_sch_result
}
\ExplSyntaxOff

\DeclareFontFamily{U}{crystallographicSymbol}{\hyphenchar\font=-1}
\DeclareFontShape{U}{crystallographicSymbol}{m}{n}{ <-> cryst}{}

\newif\ifpdffigures
\pdffigurestrue

\ifpdffigures
	\relax
\else
	\relax
	\usetikzlibrary{external}
\fi

\newcolumntype{L}[1]{>{\raggedright\let\newline\\\arraybackslash\hspace{0pt}}m{#1}}
\newcolumntype{C}[1]{>{\centering\let\newline\\\arraybackslash\hspace{0pt}}m{#1}}
\newcolumntype{R}[1]{>{\raggedleft\let\newline\\\arraybackslash\hspace{0pt}}m{#1}}

\begin{document}

\title{Soft self-assembly of Weyl materials for light and sound}

\author{Michel Fruchart}
\email{fruchart@lorentz.leidenuniv.nl}
\affiliation{Lorentz Institute, Leiden University, Leiden 2300 RA, The Netherlands}
\author{Seung-Yeol Jeon}
\author{Kahyun Hur}
\affiliation{Center for Computational Science, Korea Institute of Science and Technology, Seoul 02792, Republic of Korea}
\author{Vadim Cheianov}
\affiliation{Lorentz Institute, Leiden University, Leiden 2300 RA, The Netherlands}
\author{Ulrich Wiesner}
\affiliation{{Department of Materials Science and Engineering, Cornell University, Ithaca, New York 14850, USA.}}
\author{Vincenzo Vitelli}
\email{vitelli@uchicago.edu}
\affiliation{Lorentz Institute, Leiden University, Leiden 2300 RA, The Netherlands}
\affiliation{Department of Physics, The University of Chicago, Chicago, IL 60637, USA.}

\begin{abstract}
Soft materials can self-assemble into highly structured phases which replicate at the mesoscopic scale the symmetry of atomic crystals. As such, they offer an unparalleled platform to design mesostructured materials for light and sound. Here, we present a bottom-up approach based on self-assembly to engineer three-dimensional photonic and phononic crystals with topologically protected Weyl points. In addition to angular and frequency selectivity of their bulk optical response, Weyl materials are endowed with topological surface states, which allows for the existence of one-way channels even in the presence of time-reversal invariance. Using a combination of group-theoretical methods and numerical simulations, we identify the general symmetry constraints that a self-assembled structure has to satisfy in order to host Weyl points, and describe how to achieve such constraints using a symmetry-driven pipeline for self-assembled material design and discovery. We illustrate our general approach using block copolymer self-assembly as a model system.
\end{abstract}

\date{\today}

\maketitle

The propagation of waves in spatially periodic media is described by band theory, that determines which frequencies can propagate in a given direction and at a given wavelength. While band theory was first developed to understand the behavior of electrons and phonons in solids, it applies to all kinds of waves. For example, photonic crystals are spatially periodic structures encompassed by such a description~\cite{Joannopoulos2008}, which most notably can host a photonic band gap resulting in peculiar optical properties, such as the structural coloration of several butterflies wings \cite{Saranathan2010,Pouya2012}. 

A band structure is typically a complicated set of bands, which to a casual observer seem to cross each other every possible way. This is in fact not the case: degeneracies in a band structure mainly appear at highly symmetric points, and stem from the existence of additional symmetries (beyond translation invariance). Yet, so-called \emph{accidental degeneracies} also exist, which are not enforced by the presence of a particular symmetry \cite{Herring1937a,Herring1937b,vonNeumann1929}. Accidental degeneracies are generically rare and unstable, as they require a fine tuning of the system parameters to exist. However, such degeneracies may be protected by the existence of a particular singular configuration of the Bloch eigenstates in the neighborhood of the degenerate point (similar to a vortex or a hedgehog), which confers them a topological character and hence robustness against certain perturbations. 

The simplest of such topological degeneracies are so-called Weyl points where a three-dimensional band structure locally exhibits a linear band crossing in all directions \cite{Weyl1929,Wan2011,Armitage2017}. Crucially, such a Weyl point is characterized by a topological charge which describes the singularity in the Bloch eigenstates near the crossing point. Weyl points are robust against perturbations, which means they can be moved in momentum space, but not made to disappear unless they annihilate with a Weyl point of opposite charge, similar to hedgehog-antihedgehog pairs in real space in liquid crystals \cite{ChaikinLubensky}.
Note that the Weyl points we consider here generically occur at \emph{finite} frequency, and do not require a particular symmetry. In contrast, mechanical symmetry-protected Weyl points (similar to Dirac points in graphene) and Weyl lines occur at zero frequency \cite{Rocklin2016,Po2016,Stenull2016,Bilal2017,Baardink2017}. There, a chiral symmetry is essential to define the topological quantities, and in turn reveals a duality between zero-frequency free mechanical motions and so-called self-stress modes \cite{Kane2013,Chen2014,Paulose2015,Paulose2015b,Huber2016,Susstrunk2016}.

Excitations following the Weyl equation \cite{Weyl1929,Wan2011,Armitage2017} have been experimentally observed in electronic condensed matter in the so-called Weyl semi-metal TaAs \cite{Yang2015,Lv2015,Xu2015,Xu2015b,Lv2015b}, as well as in photonic \cite{Lu2013,Lu2015,Noh2017,Chen2016,Yang2017b,Yang2018}, phononic, and acoustic \cite{Li2017,Zhang2018,Xiao2015,Yang2016} crystals, and in homogeneous magnetized plasma \cite{Gao2016}.
Beyond their fundamental importance, such discoveries may pave the way towards multiple applications enabled by the peculiar properties of Weyl points, such as their angular and frequency selective response and the existence of topologically protected arc surface states (called Fermi arcs in the electronic context) that appear at the boundary of finite samples, even when time-reversal invariance is not broken \cite{Lu2016,Wang2016,Bi2015,Lu2016b,Hills2017,Zhou2017}. This is in sharp contrast with gapped topological materials where the existence of one-way channels requires to break time-reversal invariance in some way, such as with external drives \cite{Khanikaev2015,Wang2015,Nash2015,Fleury2016}, magnetic or rotation fields \cite{Wang2015b,Swinteck2015,Kariyado2015,Delplace2017} or active materials \cite{Souslov2017,Shankar2017}.

\begin{figure*}[htb]
\centering

\ifpdffigures
	\includegraphics{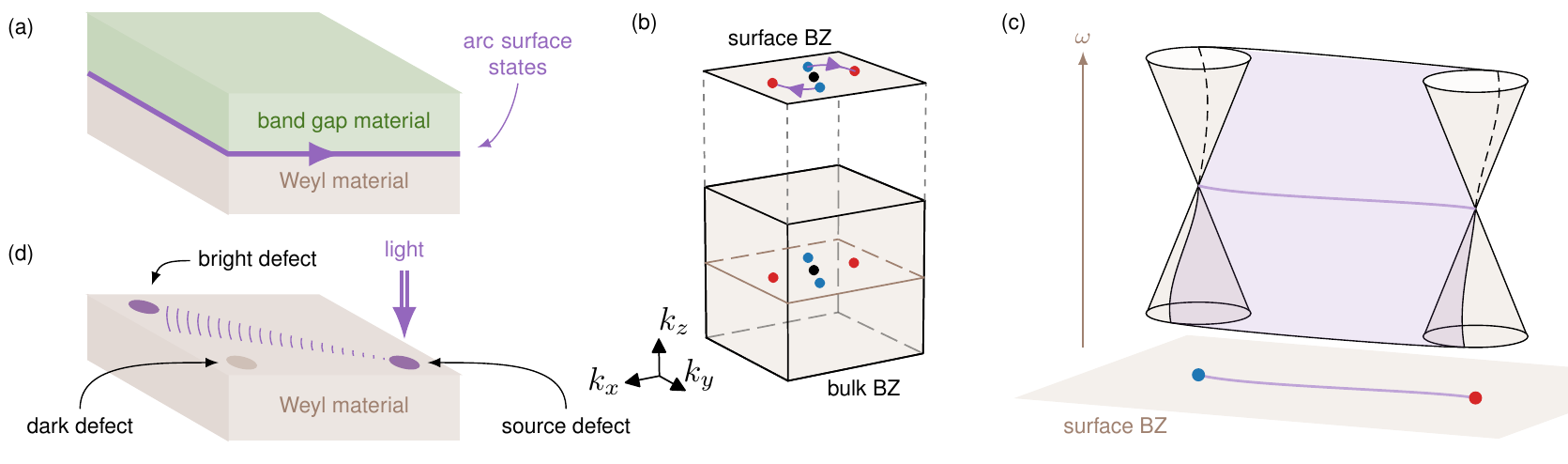}
\else
	\tikzsetnextfilename{sketch_weyl}
	\input{media/sketch_weyl.pgf}
\fi

\caption{\label{sketch_weyl}\strong{Bulk Weyl points and arc surface states.} 
(a) Sketch of an interface between a band gap material and a Weyl material. Arc surface states (in purple) appear at this interface.
(b) In a time-reversal invariant system, inversion symmetry has to be broken for Weyl points to exist. The simplest situation consists of four Weyl points with charges $\pm 1$ (respectively in red and blue) in the bulk Brillouin zone (bulk BZ). A plane interface preserves space periodicity in two directions, and is hence described by a two-dimensional surface Brillouin zone (surface BZ). Crucially, topological arc surface states (represented in purple) appear between Weyl points of opposite charge on the surface Brillouin zones.
(c) The surface dispersion relation at the interface between a Weyl material and a gapped system features conical dispersions relations, which are the projections of the Weyl points. In addition, a manifold of topological arc surface states (in light purple) appears. The intersection of this manifold with a plane of constant frequency (or energy) is sometimes called a Fermi arc, in reference to the situation in electronic solid state physics where this plane is set at the Fermi energy. (d) The arc surface states may be observed by creating defects at the interface to couple them with incident waves.
}
\end{figure*}

All photonic Weyl materials designed up to now are based on top-down approaches \cite{Lu2015,BravoAbad2015,Lin2016,Chen2016,Xiao2016,Wang2016,Noh2017,Chang2017,Wang2017,Yang2017,Yang2017b,Yang2018}. In this article, we show how soft matter self-assembly \cite{Whitesides2002} provides a viable bottom-up strategy to realize Weyl materials for sound and light. Block copolymers are used as a paradigmatic example of soft materials that self-assemble into a variety of highly structured phases arising from the competition between elastic energy and surface tension \cite{Bates1999}. However, our strategy is applicable to a wider range of self-assembled materials because it is rooted in symmetry. In the same way as the arrangement of atoms into various crystalline structures is responsible for the diverse properties of natural materials, the self-assembly of soft mesoscopic structures with various space group symmetries provides an unparalleled platform to synthesize novel materials.

Fully unleashing the potential of soft matter self-assembly in material design involves a constant interplay between the full-wave optical (or acoustic, etc.) equations of motion of the system on one hand, and its structural description in terms of free energy minimization subject to external fields and constraints on the other hand. Those problems are generally not analytically tractable, and require considerable computational power to be solved numerically for a wide range of parameters. Here, our goal is to design a bottom-up method to create Weyl materials. While self-assembly is a global process taking place in real space, Weyl points exist in reciprocal space, as they are features of the band structure describing wave propagation in the system. Hence, we have to solve an inverse problem involving both descriptions. To shortcut this difficulty, we combine a minimum input of full-wave computations with a comprehensive symmetry analysis that 
determines analytically the desired symmetry breaking fields without performing heavy numerical simulations.

This article is organized as follows. In the first section, we review the definition of a Weyl point and the properties of band structures with such singularities. This allows us to identify a first set of symmetry constraints on our candidate systems. The second section is devoted to the realization of a self-assembled block copolymer structure that meets this set of minimal requirements, namely breaking inversion symmetry. We then move on to identify what (other) symmetries should be broken to obtain Weyl points, and how to do so by applying suitable strains.
We then confirm that the designed photonic structure indeed exhibits Weyl points through full-wave computations of Maxwell equations. Simulating the self-assembled structures with broken symmetries is required to determine the \emph{quantitative} features of the band structure, and most crucially to demonstrate our method. Yet, to predict the existence of Weyl points, our framework only requires the band structure of the \emph{unmodified} system, \emph{without} symmetry-breaking alterations. This enables the extension of our design to other kinds of waves: from the full-wave band structures of the unperturbed dispersive photonic, phononic, and acoustic systems, we can predict that only the first two will exhibit Weyl points when altered and strained. The last section provides a generic blueprint for mesostructured material design by self-assembly.

\begin{figure*}[htb]

\ifpdffigures
	\includegraphics{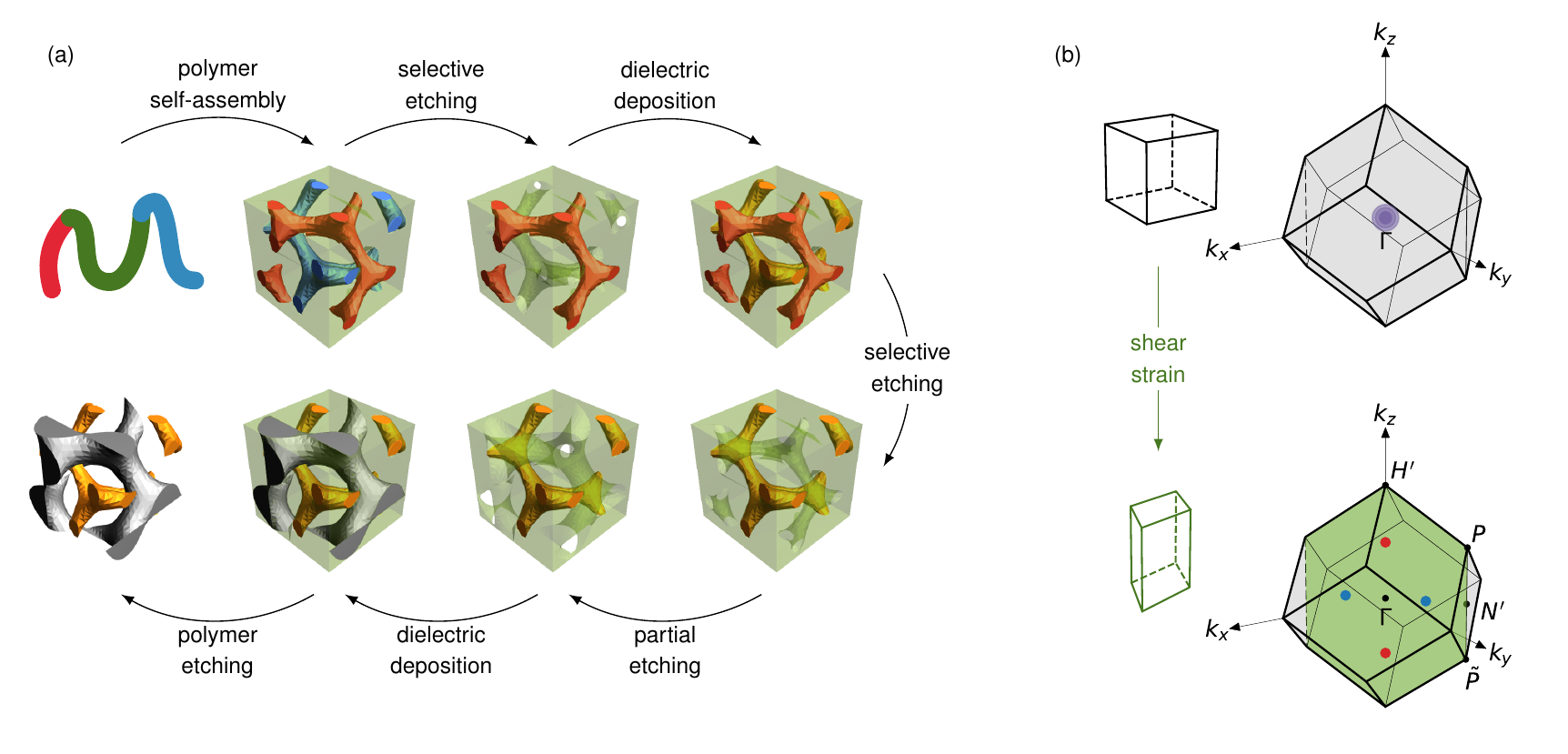}
\else
	\tikzsetnextfilename{self_assembly}
	\input{media/self_assembly.pgf}
\fi

\vspace*{-0.75cm}
  \caption{\label{self_assembly}\strong{Self-assembly process and effect of strain.}
  The self-assembly of triblock terpolymers leads to \enquote{colored} double-gyroids, where the two minority networks (in red and blue) are chemically distinct \cite{Hur2011,Cowman2015}. Starting from the self-assembled structure, a series of selective etching, partial dissolution and backfilling steps leads to an asymmetric double gyroid made of high dielectric constant materials, which constitute a three-dimensional photonic metacrystal.
  Crucially, the photonic band structure of such a system has a 3-fold degeneracy at the center of the Brillouin zone (the $\Gamma$ point), represented in purple on the top right figure. This 3-fold degeneracy can be split into a set of Weyl points by an appropriate strain (in this case, pure shear), represented in red and blue (respectively for Weyl points of charge \num{+1} and \num{-1}) on the bottom right figure.
  }
\end{figure*}

\section{Weyl points and symmetry requirements}

The three-dimensional band structure of an electronic system possessing Weyl points exhibits linear band crossings locally described by the Hamiltonian \cite{Weyl1929,Wan2011,Armitage2017}
\begin{equation}
  \label{local_weyl_hamiltonian}
  H(k) = q_i \, v_{i j} \, \sigma_{j}
\end{equation}
where $q=k-k_0$ is the wavevector relative to the Weyl point's position $k_0$, $\sigma_{j}$ are the Pauli matrices, and $v_{i j}$ is an invertible effective velocity matrix describing the band crossing at first order in $q$. While this description seems at first sight peculiar to quantum mechanical systems, it is also applicable to all kinds of waves, as we will see in the following with the example of light. Crucially, such a Weyl point is characterized by an integer-valued topological charge which describes the singularity in the eigenstates near the crossing point, and can be expressed as \cite{Wan2011,Armitage2017}
\begin{equation}
  \label{weyl_charge}
  C_1 = \text{sgn} \det (v).
\end{equation}

\begin{figure*}[htb]
\centering
\includegraphics[]{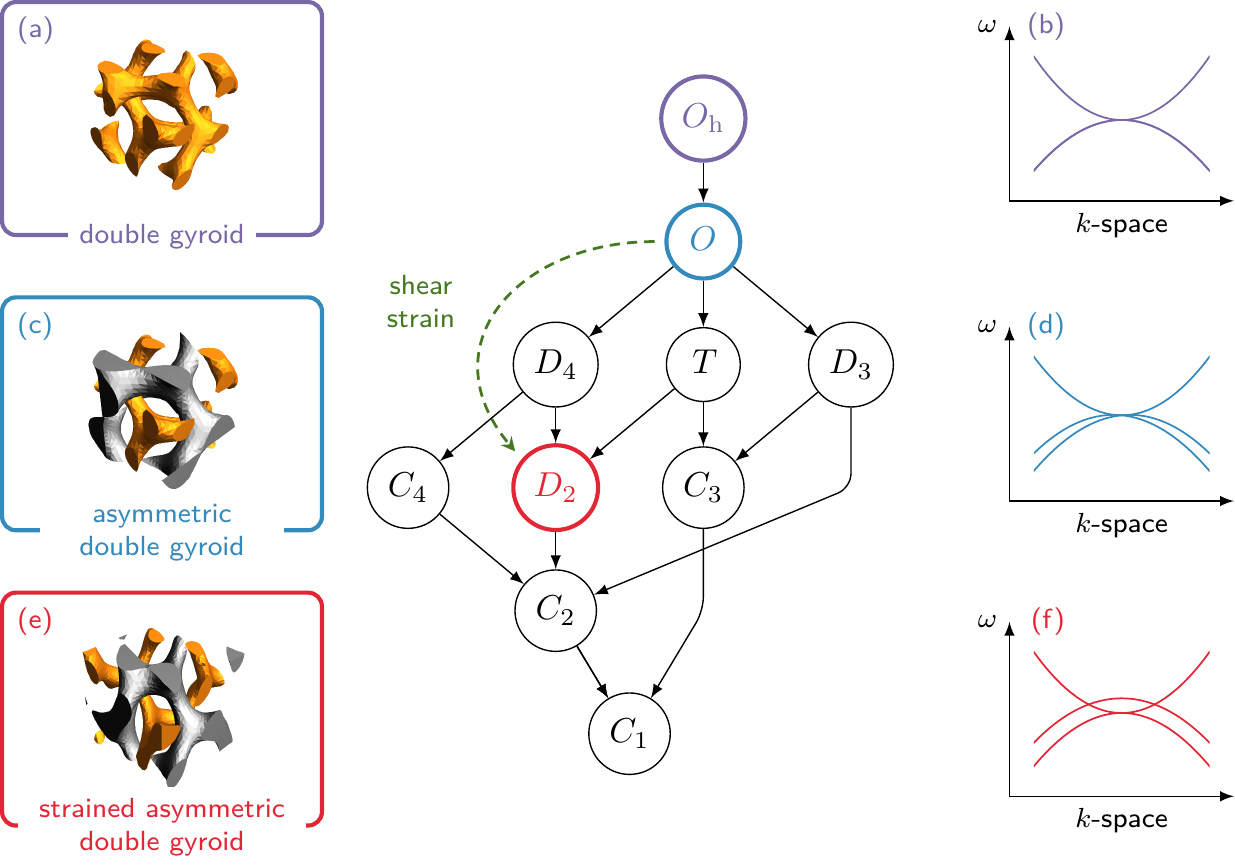}
\caption{\label{symmetry_reduction}\strong{Reducing the symmetry.}
On the left, we show the various structures of interest: (a) the (symmetric) double gyroid (DG), (b) the asymmetric double gyroid (ADG), and (c) the strained asymmetric double gyroid (with shear strain). For the double gyroid, the group of the wavevector $\Gamma$ is the octahedral group \sch{Oh} (\hm{m-3m} in Hermann-Mauguin notation), while it is the octahedral rotation group \sch{O} (\hm{432}) for the asymmetric double gyroid. 
When strain is applied to the ADG, its symmetry is reduced, which corresponds at the $\Gamma$ point to a subgroup of the octahedral rotation group \sch{O}. Such subgroups are organized in a Hasse diagram in the middle picture. 
(The octahedral rotation group is indeed a subgroup of the full octahedral group \sch{Oh}, which has more subgroup which are not relevant here, and were not represented. As strain preserves inversion symmetry, any strained symmetric double gyroid still has an inversion center. The point group at $\Gamma$ is then the product of the inversion group \sch{S2} with a subgroup of \sch{O}. Such situations can be achieved by starting from the symmetric DG. In addition, there are other mixed subgroups of \sch{Oh}, which cannot be directly realized through our method.)
On the right, the method of invariants predicts the qualitative features of the band structure of the modified double gyroids (b) and (c).
  }
\end{figure*}

Although the existence of such topologically protected Weyl points does not \emph{require} a particular symmetry, a crucial interplay between such degeneracies and symmetries exists. Notably, Weyl points cannot be obtained when both time-reversal symmetry and space inversion symmetry are present \cite{Wan2011,Halsz2012}, because inversion symmetry requires a Weyl point located at point $k$ on the Brillouin zone to have a partner of \emph{opposite} charge at $-k$, while time-reversal symmetry requires a Weyl point located at $k$ to have a partner of the \emph{same} charge at $-k$, which implies that this topological charge must be zero and no Weyl points exist.
Hence, either time-reversal or inversion symmetries (or both of them) have to be lifted to allow for Weyl points in the band structure. In a time-reversal invariant system, Weyl points come in pairs of points with identical charge and the simplest situation then consists of two such pairs with opposite charges (see figure \ref{sketch_weyl}b).

\medskip

A hallmark of Weyl materials is the existence of topological surface states at the interface with a band gap material. At a plane interface such as the one pictured in figure \ref{sketch_weyl}a, the translational invariance is preserved in two directions, and the surface is described by a two-dimensional surface Brillouin zone, as represented in figure \ref{sketch_weyl}b. In addition to conical dispersions stemming from the projection of the bulk Weyl points, the surface band structure features a manifold of \emph{arc surface states} [represented in purple in figure \ref{sketch_weyl}c] of topological origin. Let us consider a monochromatic beam of light or sound shone on the system. Depending on the wave frequency and its wavevector, it may either be reflected or transmitted in the bulk material, as ruled by the band structure. When the beam hits the interface, it may also excite the arc surface states. At all frequencies close to the Weyl points, there is an arc-shaped curve in momentum space connecting the locations of the Weyl cones, representing the set of wavevectors at which topologically protected surface states are present.

The arc surface states arise from the topology of bulk states: pictorially, they can be seen as the projections of a bulk \enquote{Dirac string} connecting the Weyl points\footnote{More precisely, the topology of the band structure is fully characterized not only by the charges of the Weyl points, but by weak first Chern numbers (or weak Fu-Kane-Mele invariants) defined on two-dimensional planes or surfaces of the bulk Brillouin zone \cite{Mathai2017a,Mathai2017b,Thiang2017}. In contrast, the exact shape of the topological surface states is indeed \emph{not} determined by the topology, and depends on the boundary conditions \cite{Yang2017b}. In particular cases, it can however be predicted from the bulk through the entire data of the Berry connection \cite{Kim2016}.}. As the Dirac string is not gauge invariant, however, topology only determines the connectivity of the surface states (i.e. which Weyl points are connected).
Beyond their fundamental importance, one of the main interest of arc surface states is the fact that their topological origin confer them a certain robustness to perturbations. Such arc states were experimentally observed in three-dimensional materials, both for light \cite{Chen2016,Noh2017,Wang2017,Yang2017b,Yang2018} and sound \cite{Xiao2015,Li2017}. Although time-reversal invariance is preserved, unidirectional wave propagation immune to backscattering can be observed at the interface \cite{Chen2016,Yang2017b,Li2017}. This robustness is however not as strong as in a system with broken time-reversal invariance, as a component of the tangent momentum has to be (at least partially) conserved \cite{Chen2016,Yang2017b,Li2017}.

While the \emph{bulk} Weyl points are most useful when they are spectrally isolated from other bands \cite{Lu2013,Lu2015}, the arc surface states do not require such a frequency isolation \cite{Yang2017b}. At microwave frequencies, arc surface states were observed by plugging an antenna into the sample \cite{Yang2017b}. At lower wavelengths, they could be observed as follows: a defect such as a small hole drilled at the surface of a Weyl material allows a coupling between the surface states and freely propagating light on the outside. Let us consider several of such holes drilled at different places on the surface of the material, as represented on figure \ref{sketch_weyl}(d). In the absence of surface states at the light's frequencies, a light beam shone on one of such holes propagates in the bulk, and quickly disappears from the interface: all the defects but the source are dark. In contrast, when surface states are present, a sizeable part of the beam intensity propagates at the interface in directions controlled by the positions of the arc states in the surface Brillouin zone. As a consequence, a handful of the holes are illuminated.
It is worth noting that the surface of a Weyl metamaterial can only support arc surface modes if the conservation laws prohibit hybridisation of such modes with the electromagnetic continuum outside. For the interface between Weyl materials and the vacuum, this requires the Weyl points to lie below the light cone in the reciprocal space. When this is not the case, one has to consider an actual interface with a band gap material. In this case, the hole can simply be extended into the band gap material.

\section{Self-assembling inversion-asymmetric gyroids}
\label{self_assembly_inversion_asymmetric}

In this article, we assume time-reversal invariance and concentrate on inversion symmetry breaking to avoid the need of external drives, magnetic fields or active materials. However, generically breaking inversion symmetry leads to uninteresting band structures. We adopt the following strategy: (i) start with highly symmetric structures possessing particular degeneracies (ii) split such degeneracies into Weyl points by applying carefully chosen symmetry-breaking perturbations. Implementing both steps through a bottom-up strategy is very challenging. Our goal is to overcome this difficulty using soft self--assembly. 

\medskip

The first example of a photonic crystal displaying Weyl points was engineered by milling and stacking dielectric layers into a highly symmetric structure called a \emph{double gyroid}, in which additional holes were deliberately drilled at strategic points to reduce the symmetry of the system \cite{Lu2015}.
A gyroid is an infinitely connected triply periodic surface of zero mean curvature, discovered by Alan Schoen \cite{Schoen1970}. The surface of the gyroid divides space into two regions corresponding the interpenetrating labyrinth structures shown in Fig. \ref{self_assembly}.

A remarkable fact from soft matter science is that double gyroids naturally self-assemble in situations where two or several linked components have repulsive interactions with each other. In such circumstances, the minimization of surface energy constrained by the presence of links between the immiscible components can lead to a variety of \emph{minimal surfaces}, among which is the gyroid surface. Gyroids generically appear in various soft materials such as liquid crystals \cite{Longley1983,Mezzenga2005,Aplinc2016}, amphiphilic surfactants \cite{Fontell1990,Monnier1993}, dispersions of anisotropic and patchy colloids \cite{Glotzer2007,Sacanna2010,Wang2012,Phillips2012,Marson2014}, and block copolymers \cite{Schulz1994,Hajduk1994,Bates1999} to name but a few.

\medskip

AB diblock copolymers are the archetypal example of such a self-assembling soft material. They are composed of two immiscible polymer blocks denoted by A and B, glued-together by covalent bonds. For a well-chosen set of the system parameters (typically the average degree of polymerization, the relative fractions of A and B, and the Flory-Huggins parameter characterizing the interaction energy between the block A and B), the constrained minimization of the interface energy leads to a double gyroid structure, where two minority networks of opposite chirality are interwoven inside a matrix majority network \cite{Matsen1998,Meuler2009}. The interface between one of the minority networks and the matrix is a \emph{gyroid surface}, a triply periodic constant curvature surface \cite{Schoen1970,Karcher1989,GrosseBrauckmann1996}, which is well approximated by the isosurface $g(x,y,z) \equiv \sin(2 \pi x) \cos(2 \pi y) + \sin(2 \pi y) \cos(2 \pi z) + \sin(2 \pi z) \cos(2 \pi x) = t$ [with $0 \leq t < \sqrt{2}$] \cite{Wohlgemuth2001}, where $x$, $y$, and $z$ are measured in units of the unit cell size $a$. The second minority network is obtained from the first through space inversion. Hence, one of the gyroidal minority networks is described by $g(x,y,z) \geq t$  while its chiral partner, obtained by space inversion, is described by $g(-x,-y,-z) \geq t$. Both are composed, say, of the A blocks, while the majority matrix is composed of the B blocks. 

Crucially, the resulting structure has inversion symmetry, that is almost impossible to get rid of without local modifications. 
This is certainly possible in engineered structures like the milled structures of Ref. \cite{Lu2015} where one has direct control on the shape of the unit cell but it is not compatible with  
a bottom-up material synthesis scheme. In order to take advantage of a self-assembly scheme, we instead choose to use ABC triblock terpolymers which self-assemble in a double-gyroid where two \emph{chemically distinct} gyroid-shaped minority networks of opposite chirality are interwoven inside a matrix majority network \cite{Matsen1998,Meuler2009}. For instance, one of the gyroidal labyrinths may be composed of A blocks, but its image by space inversion is then composed of C blocks, while the matrix is still composed of B blocks. The resulting structure is called an \emph{asymmetric double gyroid} or an \emph{alternating double gyroid}.
After the polymer self-assembly, standard techniques allow to selectively etch one of the gyroidal minority network and to replace it with a high-permittivity material \cite{Park2003,Yoon2005}, for example by metal \cite{Hur2011,Cowman2015} or dielectric \cite{Fink1999,Urbas2002} deposition. Crucially, the chemical difference between both gyroidal networks allows to induce an optical asymmetry between them, either by depositing materials of different dielectric constants or through the use of a mild etching agent to tune their respective radii. The last step is to get rid of the majority network matrix. The whole process is summarized pictorially in figure~\ref{self_assembly}. After this process is complete, we are left with a structure where the dielectric constant is $\varepsilon_{A}$ for $g(x,y,z) \geq t_A$, $\varepsilon_{B}$ for $g(-x,-y,-z) \geq t_B$, and is $\varepsilon_{\text{air}} = 1$ outside of such regions.
Similarly, a phononic crystal can be obtained by inducing an asymmetry in the elastic properties of the two gyroidal networks. In the following, we focus on photonic crystals for concreteness, but full details on acoustic and phononic crystals are provided in the Supporting Informations.

\section{Effective description of the band structure}

In order to obtain Weyl points, the symmetry of the double gyroid must be reduced further.
Full-wave numerical simulations reveal that the photonic band structure of a dielectric double gyroid has a three-fold quadratic degeneracy at the $\Gamma$ point [the center of the first Brillouin zone] \cite{Maldovan2002}. From the point of view of symmetries, the three-fold degeneracy is allowed by the existence of three-dimensional irreducible representations of the subgroup of symmetries leaving the $\Gamma$ point invariant, namely the irreducible representation \irrep{T1g} of the full octahedral group \hm{m-3m} (or \sch{Oh} in Schoenflies notation) [see Supporting Informations]. This three-fold degeneracy can be split into pairs of Weyl points by symmetry-breaking perturbations \cite{Lu2013,Wang2016}, as represented in figures \ref{self_assembly}b and \ref{symmetry_reduction}. The systematic description of a band structure near a high-symmetry point of the Brillouin zone as well as the effect of symmetry-breaking perturbations can be obtained from group theory. This approach known as method of invariants originated within condensed matter physics \cite{Luttinger1956,Pikus1961,BirPikus1975,Winkler2003,Willatzen2009}, but it also applies to photonic systems \cite{Sakoda1995,Sakoda2004} and more generally to all kinds of waves in periodic media. 

For example, in the absence of charges and currents Maxwell equations can be written in a convenient way as
\begin{equation}
  \label{maxwell_schrodinger}
  \ii \partial_t
  \begin{pmatrix}
    \mathcal{E} \\ \mathcal{H}
  \end{pmatrix}
  =
  \left[
  \begin{pmatrix}
    \varepsilon & 0 \\
    0 & \mu 
  \end{pmatrix}^{-1}
  \begin{pmatrix}
    0 & \ii \, \text{rot} \\
    - \ii \, \text{rot} & 0 \\
  \end{pmatrix}
  \right]
  \begin{pmatrix}
    \mathcal{E} \\ \mathcal{H}
  \end{pmatrix}
\end{equation}
where $\mathcal{E}$ and $\mathcal{H}$ are the electric and magnetic fields, while $\epsilon$ and $\mu$ are the spatially varying permittivity and permeability of the medium .
In this form, the operator in square brackets, called the Maxwell operator, plays the role of a Hamiltonian\footnote{For normal materials where permittivity and permeability are strictly positive, the Maxwell operator is Hermitian with respect to a relevant scalar product \cite{Joannopoulos2008,DeNittis2014}. An additional constraint stemming from the source-free equations has to be taken into account, which however commutes with the Maxwell operator.} [see e.g. \cite{Joannopoulos2008,DeNittis2014,DeNittis2017}]. This full-wave Maxwell equation is usually impossible to solve analytically: one has to resort to numerical simulations. However, with a minimal input from a full-wave solution complemented with the full knowledge of symmetries in the problem, one can determine an \emph{effective} Hamiltonian which is sufficient for perturbative design purposes. Similar considerations apply to other kinds of waves propagating in periodic media (for example elastic waves, see Supporting Informations).

By reducing the full description of the system (contained in the Maxwell operator) to the subspace spanned by a few relevant degrees of freedom, one obtains an \emph{effective Hamiltonian} describing a few bands in the vicinity of a (usually high-symmetry) point $k_0$ of the Brillouin zone. For example, the eigenstates involved in a degeneracy at $k_0$ can then be used as a basis to describe the effective Hamiltonian operator as a matrix $H(q)$ where $q=k-k_0$. 
Both the Maxwell operator and the effective Hamiltonian operator are invariant with respect to the symmetries $g$ of the group of the wavevector $k_0$, defined as the subgroup of symmetries which leave $k_0$ invariant. 
The general idea of the theory of invariants is that symmetries can be used to construct the effective Hamiltonian matrix from scratch, by combining a set of \emph{basis matrices} $X$ (which form a basis of, say, the space of $3 \times 3$ Hermitian matrices) and \emph{irreducible functions} $\mathcal{K}(q)$ of the wavevector components (like $q_x^2 + q_y^2 + q_z^2$). The basis matrices represent operators in the basis of eigenstates at $k_0$. As such, they change when a symmetry operation is applied. This is also the case of irreducible functions as the symmetry operation is applied to the momentum vector. As the effective Hamiltonian operator is \emph{invariant} when a symmetry is applied, it is possible to determine all terms allowed by symmetry by selecting all combinations of the form $\mathcal{K}(q) X$ which are left invariant by the action of the symmetries.

More precisely, if the eigenstates at $k_0$ form an irreducible representation $\Gamma$, then the matrix representation $H(q)$ of the effective Hamiltonian operator describing the corresponding bands will be covariant with respect to the symmetries. Namely, $D(g) H(g^{-1} q) D(g)^{-1} = H(q)$, where $D$ is a representation of $\Gamma$, acting on the effective Hamiltonian by its adjoint action as a representation of $\Gamma \times \Gamma^{*}$. The effective Hamiltonian can then be constructed by combining basis matrices $X^{(\gamma,\mu)}$ of the irreducible representations $\Gamma_{\gamma}$ appearing in decomposition of this product and irreducible functions of the wavevector components $\mathcal{K}^{\gamma,\mu}(q)$ as
\begin{equation}
  \label{eq:sum_invariants}
  H(q) = \sum_{\gamma} a_{\gamma} \sum_{\mu} X^{(\gamma,\mu)} \mathcal{K}^{\gamma,\mu *}(q)
\end{equation}
where $a_{\gamma}$ are arbitrary constants chosen such that $H(q)$ is Hermitian, and where $\mu$ labels the basis elements in a same irreducible representation.

Given the input of (a) the space group \hm{Ia-3d} of the double gyroid and (b) the fact, known from full-wave computations, that a three-fold band crossing transforming according to the three-dimensional irreducible representation \irrep{T1g} exists at the $\Gamma$ point, the method of invariants yields the following effective Hamiltonian, describing the band structure in the vicinity of this crossing (see Supporting Informations)

\begin{equation}
\begin{split}
  \label{eq:Luttinger_Hamiltonian}
  H_{0}(k) = & \; a_{1}^{(0)} \; \Id
  + a_{1}^{(2)} \; k^2 \; \Id
  + a_{12}^{(2)} \; \big(
    \overline{K} \; \Lambda + 
    K \; \Lambda^{\dagger}
  \big) \\
  + &\; a_{25'}^{(2)} \; \big(
    k_x k_y \; L_x L_y + \text{c.p.}
  \big) + O(k^3)
\end{split}
\end{equation}
\noindent where $k^2 = k_x^2 + k_y^2 + k_z^2$, and \enquote{c.p.} stands for \enquote{circular permutation} (of the indices). Matrices $L_i$ are three-by-three Hermitian angular momentum matrices satisfying $[L_i,L_j] = \ii \, \epsilon_{i j k} L_k$ and $L_x^2 + L_y^2 + L_z^2 = 2 \Id$, and $\Id$ is the identity matrix.
We also defined $K = (k_x^2 + \overline{\omega} k_y^2  + \omega k_z^2)$ and $\Lambda = L_x^2 + \overline{\omega} L_y^2 + \omega L_z^2$ with $\omega = \ee^{\ii 2\pi/3}$.
In this expansion, the indices of the coefficients $a_{I}^{(p)}$ refer to the irreducible representation $\Gamma_{I}$ from which the invariant term was constructed from, and the exponent in parentheses to the order of the irreducible polynomial composed of the wavevector components (at the $\Gamma$ point, $k_0=0$, so $q=k$). Finally, in addition to space symmetries, time-reversal invariance is imposed by considering only time-reversal-even combinations.

The main interest of the method of invariants is that it allows to determine what \emph{new} terms can be added to the preceding effective Hamiltonian when the symmetry is reduced. This enables to qualitatively predict the effect of perturbations on the band structure, as illustrated in figure \ref{symmetry_reduction}. As we have seen, the very first step towards inducing Weyl points is the removal of inversion symmetry from the structure (see figure \ref{symmetry_reduction}c). Hence, the point group at $\Gamma$ becomes the chiral octahedral group \hm{432} (or \sch{O} in Schoenflies notation). However, this modification does not allow a constant term in the effective Hamiltonian: only a new linear term of the form $k_x L_x + k_y L_y + k_z L_z$ appears due to the reduction of the irreducible representation $\text{T}_{1\text{g}}$ and $\text{T}_{1\text{u}}$ of \sch{Oh} to $\text{T}_{1}$ in \sch{O}. Hence, the quadratic band crossing at $\Gamma$ cannot be lifted by such a term (see figure \ref{symmetry_reduction}d), and a further reduction in symmetry is required.

\begin{figure}[htb]
\centering
\hspace*{-0.5cm}

\ifpdffigures
	\includegraphics{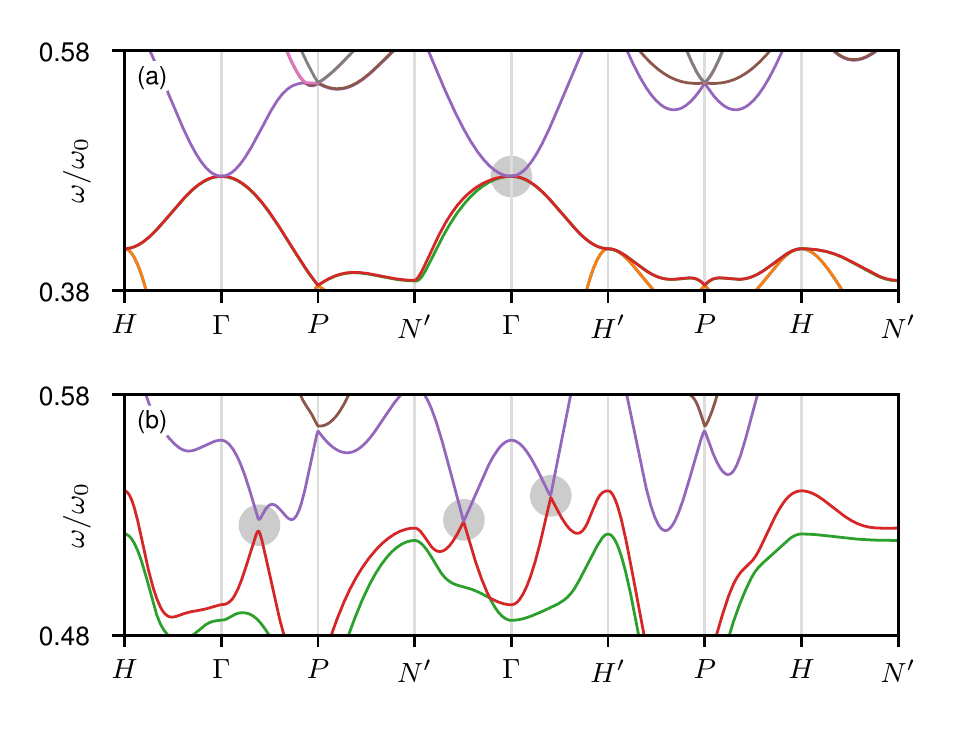}
\else
	\tikzsetnextfilename{band_structures}
	\input{media/band_structures.pgf}
\fi

\vspace*{-1cm}
\caption{\label{band_structures}\strong{Photonic band structures.}
Photonic band structures of (a) the symmetric double gyroid and (b) the shear-strained asymmetric double gyroid. 
The 3-fold quadratic band crossing at the $\Gamma$ point of the band structure of the unperturbed double gyroid is split into Weyl points on the $\Gamma-N'$ and $\Gamma-H'$ lines (in constrast, there is no crossing on the $\Gamma-P$ line, which distinguishes the pair of Weyl points from a nodal line). 
The first $8$ bands of the band structures were computed with the MPB package \cite{Johnson2001} on a $(64 \times 3)^{3}$ grid, with
(a) $\varepsilon_{A} = \varepsilon_{B} = \num{16}$, $t_{A} = t_{B} = \num{1.1}$, and $\theta=0$ ;
(b) $\varepsilon_{A} = \num{20.5}$, $\varepsilon_{B} = \num{11.5}$, $t_{A} = t_{B} = \num{1.1}$, and $\theta=0.3$.
Here, $\omega_{0} = 2 \pi c/a$ where $c$ is the speed of light in vacuum.
}
\end{figure}

\begin{figure}[htb]
\centering
\hspace*{-0.5cm}

\ifpdffigures
	\includegraphics{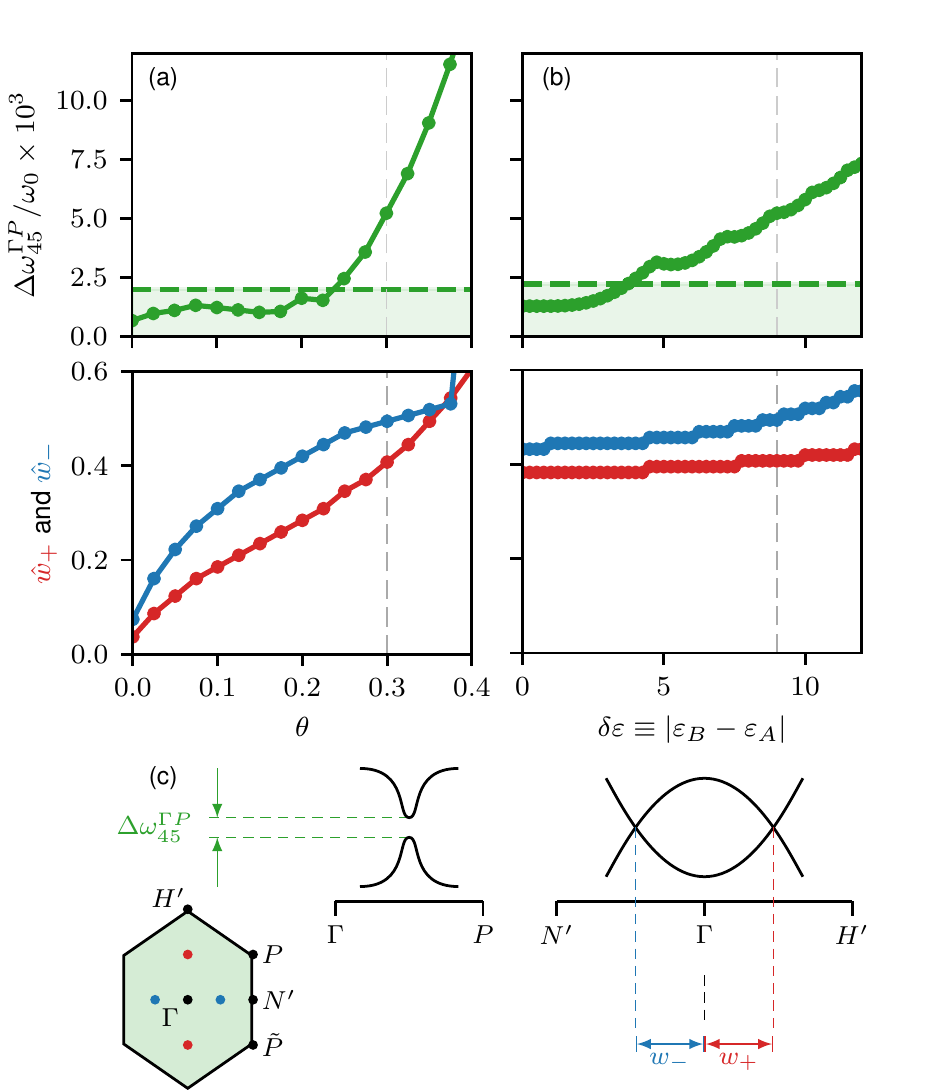}
\else
	\tikzsetnextfilename{main_features}
	\input{media/main_features.pgf}
\fi

\vspace*{-0.5cm}
\caption{\label{main_features}\strong{Evolution of the main features of the photonic band structures.}
Evolution of characteristic features of the Weyl points with the strain angle $\theta$ and the dielectric asymmetry $\delta\varepsilon \equiv \varepsilon_{B} - \varepsilon_{A}$. Here, the band structures are computed with the set of parameters (b) of figure \ref{band_structures} on a $(32 \times 3)^{3}$ grid.
We plot both the minimum of the local gap between the $4^{\text{th}}$ and the $5^{\text{th}}$ bands, $\Delta f_{4 5}^{\Gamma P} = \min \{ | f_5(k) - f_4(k) | \mid k \in \Gamma-P\}$, and the normalized positions of the two inequivalent Weyl points, $\hat{w}_{+} = w_{+}/\lVert \Gamma N'\rVert$ and $\hat{w}_{-} = w_{-}/\lVert \Gamma H'\rVert$.
The gray dashed lines correspond to the value at which each parameter is kept constant in other figures.
The abrupt jump in the position of one of the Weyl points near $\theta=\num{0.4}$ is an artifact: another set of band crossings appears on the $\Gamma-N'$ and $\Gamma-H'$ lines near this value (see Supporting Informations). 
For the local gap, a light green region delimited by a dashed line corresponds to the order of magnitude of symmetry-breaking numerical errors (see Supporting Informations). The data is not meaningless below this threshold, but the effect of the strain and structural asymmetry are not distinguishable from the spurious numerical reduction of symmetry. Similarly, $\hat{w}_{\pm}$ should both vanish at $\theta=0$ (which is clearly not the case). This provides an order of magnitude of the uncertainty on both observables.
}
\end{figure}

\section{Strain and symmetries}
\label{strain_symmetries}

The simplest yet \emph{global} way to reduce the symmetry of a structure is to apply a mechanical strain. The key point is that this strategy is compatible with self-assembly, unlike local modification or patterning of the individual building blocks. The asymmetry between the enantiomeric (i.e. non-identical mirror images of each other) gyroidal components described in the previous section reduces the space group \hm{Ia-3d} (IUC 230) of the symmetric double gyroid to \hm{I4_132} (IUC 214). When strain is applied, this space group is further reduced.

We choose to apply the shear strain
\begin{equation}
  \label{shear_strain}
  \epsilon = \begin{pmatrix}
    \cos\theta & \sin \theta & 0 \\
    \sin \theta & \cos \theta & 0 \\
    0 & 0 & 1 \\
  \end{pmatrix}
\end{equation}
written in the cartesian coordinates of the standard conventional cell (not the primitive cell). This transformation, illustrated in figure \ref{symmetry_reduction}e), reduces the space group of the asymmetric double gyroid to \hm{F222} (IUC 22) when $\theta$ is non-zero (the method to compute the space groups of the structures, based on the open-source spglib library \cite{spglib}, is detailed in the Supporting Informations). Correspondingly, the point group at the $\Gamma$ point is \hm{222} (or \sch{D2} in Schoenflies notation). As we shall see, the effective description of the band structure near the $\Gamma$ point predicts the appearance of Weyl points in this situation.

\medskip 

The effect of a reduction in symmetry on the effective Hamiltonian can be determined using subduction rules between the original symmetry group and its subgroup, which describe how the original irreducible representations combine into the new ones. In a system with lower symmetry, it is possible to combine some $X^{\gamma}$ and $\mathcal{K}^{\delta}$ in a way which was previously not allowed. In our case, going from \sch{Oh} to
\sch{D2} allows various new terms in the effective Hamiltonian, which becomes $H(k) = H_{0}(k) + \Delta H(k)$ where
\begin{equation}
  \label{new_terms_effective_hamiltonian}
  \Delta H(k) = \beta_{\pm} \; \Lambda_{\pm} + \gamma_{i} \; k_i \, L_i + \delta_{\pm} \; K_{\pm} \, \Id + \zeta_{\pm} \; k^2 \, \Lambda_{\pm}
\end{equation}
where $\Lambda_{+} = \Lambda + \Lambda^{\dagger}$ and $\Lambda_{-} = \ii (\Lambda - \Lambda^{\dagger})$ while $K_{+} = \overline{K}$ and $K_{-} = K$. Implicit summation over $i=x,y,z$ and $\pm$ is assumed. (We imposed, as an additional constraint, that time-reversal symmetry be preserved.) In this expression, $\beta_{\pm}$, $\gamma_{i}$, $\delta_{\pm}$, $\zeta_{\pm}$ are generically non-vanishing free parameters, which depend on the details of the system.
The strained \emph{symmetric} double gyroid can also be described in such a way: the only difference is that all symmetry groups now include inversion symmetry. Shear strain then reduces the point group at $\Gamma$ from \sch{Oh} to \sch{D2h}, which imposes $\gamma_i = 0$. 

Particularly noteworthy in \eqref{new_terms_effective_hamiltonian} are the constant terms with prefactors $\beta_{\pm}$ which allow the three-fold degeneracy at $\Gamma$ to be lifted. As such constant terms do not break inversion symmetry, they cannot single-handedly lead to the appearance of Weyl points. Instead, they split the three-fold degeneracy into an entire nodal line of degeneracies, similar to the one predicted in \cite{Lu2013}, which is robust against (small) inversion-preserving perturbations.
In contrast, a perturbation of the form \eqref{new_terms_effective_hamiltonian} generically breaks inversion symmetry and produces Weyl points, as observed in figure \ref{symmetry_reduction}f (see also Supporting Informations for typical spectra of the effective Hamiltonians with different symmetries). Hence, we can predict that in a well-chosen parameter range, the strained asymmetric double gyroid will exhibit Weyl points.

\section{Numerical computation of photonic band structures}

To confirm the existence of Weyl points in the strained asymmetric double gyroid, we proceed to a full-wave computation of the band structure using the well-established open-source package MPB, which determines the fully-vectorial eigenmodes of Maxwell equations with periodic boundary conditions \cite{Johnson2001}. Linear crossings between the $4^{\text{th}}$ and $5^{\text{th}}$ bands are observed in the situation described in the previous section, see figure \ref{band_structures}b (the relevant bands are red and purple, and the Weyl points and avoided crossing are marked by gray disks). In this case, the difference between a nodal line and a set of Weyl points has to be searched on the $\Gamma-P$ line. In the asymmetric double gyroid structure, a local gap separates the $4^{\text{th}}$ and $5^{\text{th}}$ bands along this line, which closes in the inversion-symmetric structure. 
To ensure that such crossings are indeed Weyl points, we compute their topological charge from the numerically computed eigenmodes (see Supporting Informations).
We find that the topological charge of the crossing point on the $\Gamma-H'$ axis is $+1$, while the charge of the crossing point on the $\Gamma-N'$ axis is $-1$ (the crossing points on the $\Gamma-\overline{N'}$ and $\Gamma-\overline{H'}$ axes have the same charge as their time-reversal counterparts).

\medskip

Either an asymmetry only in the dielectric constants or in the gyroids' thicknesses is sufficient to obtain photonic Weyl points. The effects of both perturbations are similar, but not identical: their combination may allow to optimize for additional features in the band structure (not necessarily topological), for example avoiding frequency overlaps (see Supporting Informations for an example). Here, we focus on the Weyl points: the effect of the dielectric constant asymmetry on the local gap on the $\Gamma-P$ line and on the positions of the Weyl points is shown in figure \ref{main_features}b (the effect of the gyroid thickness asymmetry can be found in the Supporting Informations). While the strain angle affects both the relative positions of the Weyl points and the gap on $\Gamma-P$ (figure \ref{main_features}a), the relative positions of the Weyl points are almost not affected by the asymmetry.

Additionally, both the dielectric and thickness asymmetries gradually open a complete band gap between the $2^{\text{nd}}$ and $3^{\text{rd}}$ bands. Here, this effect is unwanted, as it reduces the bandwidth available for the Weyl points. It may however turn out to be useful in other contexts. As the strain tends to reduce the size of this band gap, we also obtain a three-dimensional strain-tunable photonic band gap material \cite{Kim2001,Li2004}, whose properties can be adjusted through the dielectric and thickness asymmetries. Such tunable gap materials have been used to realize strain sensors \cite{Fortes2011}. Here, we can envision a combination of such strain-sensing methods with an optical tracking of the strain-induced Weyl points to achieve a high-precision measure of mechanical properties.

\begin{figure}[htb]
    \centering
    \hspace{-1cm}\includegraphics{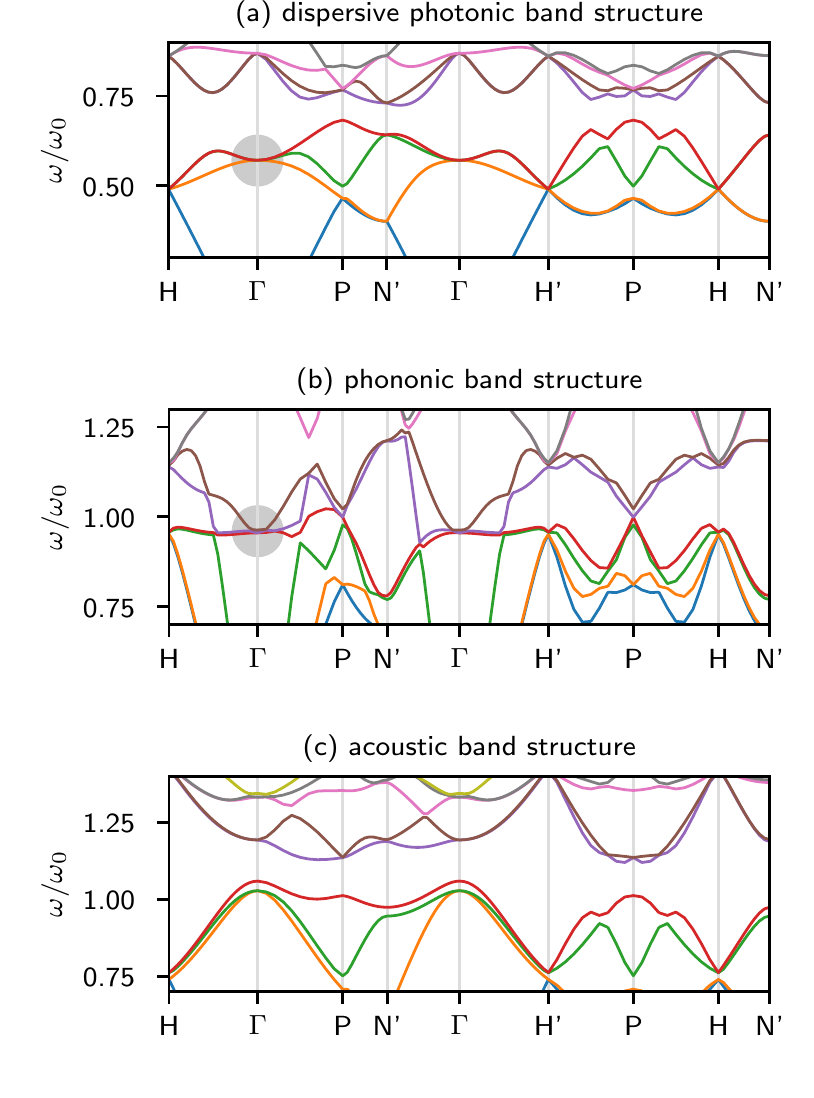}
    \vspace{-0.5cm}
    \caption{\label{different_waves}\strong{Band structures of the unperturbed double gyroid for different waves}
    (a) Dispersive photonic band structure of a metallic double gyroid structure made of a Drude metal with the plasma frequency of gold standing in vacuum. (b) Phononic band structure for an elastic double gyroid in steel embedded in an epoxy elastic matrix. (c) Acoustic band structure for sound in air confined outside of a double gyroid, with hard wall boundary conditions.
    In the dispersive photonic and phononic band structure (a) and (b), a three-fold degeneracy (highlighted by a gray circle) is found. As such, we expect such systems to exhibit Weyl points when strained.
    In (a), $\omega_{0} = 2 \pi c/a$ where $c$ is the speed of light in vacuum. We use the plasma frequency of gold, $\omega_{\text{p}}/2\pi \simeq \SI{2.19e15}{\hertz}$ \cite{Ordal1983} and $a \simeq \SI{500}{\nano\meter}$. The loss term $\Gamma$ is initially set to zero, and the results show no significant deviations from the case computed with the tabulated value $\Gamma/2\pi=\SI{5.79e12}{\hertz}$ \cite{Ordal1983}.
    In (b) $\omega_{0} = 2 \pi c_{\text{t}}/a$ where $c_{\text{t}}$ is the speed of transverse waves in epoxy. The values assumed for the longitudinal and transverse speeds of sound in steel and epoxy are obtained from the components of elastic tensor $C_{I J}$ as $c_{\text{t}}^2 = C_{4 4}/\rho$ and $c_{\ell}^2 = C_{1 1}/\rho$ from the values of references \cite{Vasseur2001,Sun2005,Hsieh2006}, namely
    $\rho^{\text{epoxy}} = \SI{1180}{\kilogram\per\meter\cubed}$, $C_{1 1}^{\text{epoxy}} =  \SI{7.61}{\giga\pascal}$, $C_{4 4}^{\text{epoxy}} =  \SI{1.59}{\giga\pascal}$ and $\rho^{\text{steel}} = \SI{7780}{\kilogram\per\meter\cubed}$, $C_{1 1}^{\text{steel}} =  \SI{264}{\giga\pascal}$, $C_{4 4}^{\text{steel}} =  \SI{81}{\giga\pascal}$.
    In (c) $\omega_{0} = 2 \pi c_{\text{air}}/a$ where $c_{\text{air}}$ is the speed of sound in air.
    All computations are performed with a $48 \times 48 \times 48$ grid. For more details on the model and computation, see Supporting Materials.
}
\end{figure}

\section{Self-assembled Weyl materials for light and sound}

While we focused on photonic systems, the same group-theoretical analysis applies to other kinds of waves. We consider three examples, where Weyl points were already demonstrated: (a) dispersive photonic media \cite{Gao2016} [i.e. with a frequency-dependent dielectric tensor] (b) phononic crystals \cite{Zhang2018} and (c) acoustic crystals \cite{Xiao2015,Yang2016,Li2017}.
To demonstrate the possibility of obtaining a \emph{self-assembled} Weyl material, we only need to consider the band structure of an unperturbed double gyroid, and look for an essential three-fold degeneracy at the $\Gamma$ point. The rest follows from our group-theoretical analysis. As we shall see, such a three-fold degeneracy appears both in dispersive photonic and phononic systems, but does not seem to arise in the considered acoustic system, at least at reasonably low frequencies. 

When light propagates in a structure made of a metal or in a dielectric at high frequency, the plasma oscillations of the electron density couple with the electromagnetic field, leading to a \emph{dispersive} photonic crystal where the propagation of light is still described by Maxwell equations, but with a frequency-dependent dielectric tensor \cite{Raman2010,AshcroftMermin}. We consider a double gyroid made of a Drude metal with the plasma frequency of gold, the band structure of which is represented in figure \ref{different_waves}a. In dispersive photonic crystals, the scale invariance of Maxwell equations is not valid anymore, as the plasma frequency provides a length scale. The case of a unit cell of size $a=\SI{100}{\nano\meter}$ was previously considered in \cite{Hur2011}. At such scales, the three-fold degeneracy may still be present, but is overlapped by highly-dense plasma bands and cannot be identified. Here, we consider a unit cell of size $a=\SI{500}{\nano\meter}$.  A three-fold degeneracy appears near $\omega/\omega_{0} \sim \num{0.5}$. Inspection of the eigenvectors shows that the electric field transforms along the three-dimensional irreducible representations \irrep{T1u} (see Supporting Informations for more details). Similarly, in a phononic crystal, elastic waves propagate in a spatially periodic structure. Here, we consider a double-gyroid made of steel embedded in an epoxy matrix, which couples elastically the two enantiomeric gyroids [see also \cite{Hur2017}]. As observed in figure \ref{different_waves}b, a three-fold degeneracy is found near $\omega/\omega_{0} \sim \num{0.95}$. Inspection of the numerical eigenvectors shows that they also transform according to the three-dimensional irreducible representation \irrep{T1u} (see Supporting Informations for more details).  According to our analysis, such three-fold degeneracies will be split into Weyl points by inducing an asymmetry in the enantiomeric gyroid networks and applying an appropriate strain, both in the dispersive photonic and phononic systems.

By contrast, we consider the case of an acoustic system, where sound propagates in air outside a double-gyroid-shaped labyrinth. Here, no three-fold degeneracy at $\Gamma$ seems to appear in the band structure [at least below $\omega/\omega_{0} \sim \num{1.75}$, see figure \ref{different_waves}c and Supporting Informations], for the values of the parameters we considered. As a consequence, we do not expect Weyl points to appear under strain at those frequencies.
Finally, the band structure of electrons constrained to move on gyroid-shaped nanostructured semiconductors displays multiple degeneracies \cite{Koshino2005,Khlebnikov2009}, which could also give rise to Weyl points under strain. In this situation however, one has to take into account the spin degrees of freedom of the electrons, which are also affected by the curvature, so we can draw no definitive conclusion from our analysis, which would have to be adapted to include double group representations.

\def\topfraction{0.99}

\begin{figure}[htb]
    \centering
    
\ifpdffigures
	\includegraphics{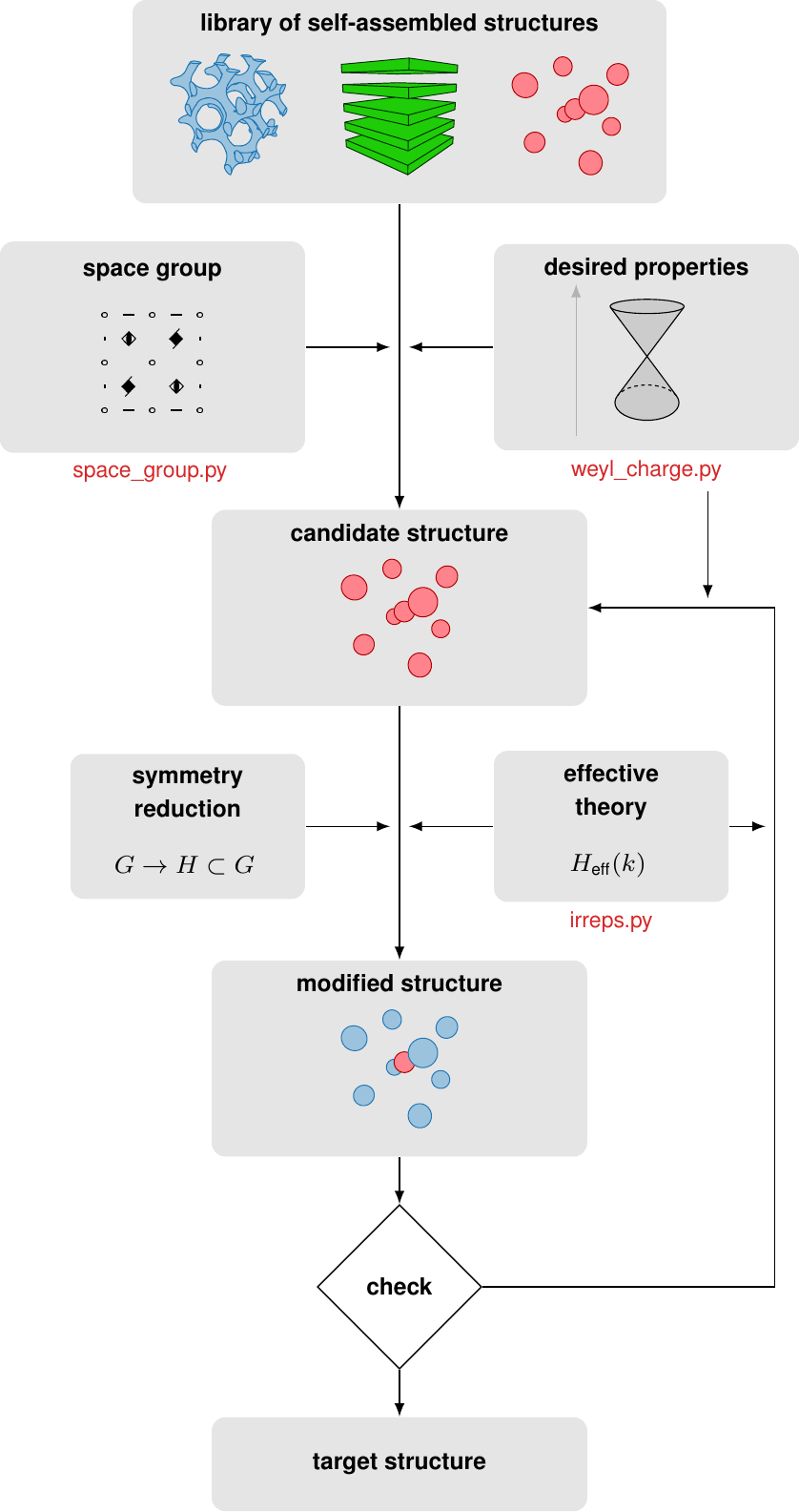}
\else
	\tikzsetnextfilename{summary_flowchart}
	\input{media/summary_flowchart.pgf}
\fi

    \caption{\label{summary_flowchart}\strong{Symmetry-driven mesostructured material discovery pipeline.} To obtain mesostructured materials with a set of desired properties, we suggest the following automated discovery pipeline. We start from a library of self-assembled structures, which is scanned for candidates matching symmetry requirements for a set of target properties. This requires to automatically determine the space group of each structure: a script space\_group.py does this job for structures represented as a skeletal graph (see Supporting Informations). A best candidate for the initial structure is then selected, and its properties are numerically computed. For example, we compute the band structure, from which the topological charges of the Weyl points (if any) are determined by a script weyl\_charge.py. An effective description is then extracted from the numerical data: here, we need to determine the irreducible representations of the numerical eigenvectors, a job performed by the script irreps.py (see Supporting Informations). The effective description then allows to determine which modifications should lead to the desired properties, for example through a symmetry reduction. This step could also be automated using \protect\url{https://github.com/greschd/kdotp-symmetry}.
    Finally, the properties of the modified structure are numerically determined, and compared against the desired properties. In case of failure, a new initial structure is selected from the library and the process is iterated.}
\end{figure}

\medskip

The self-assembly of asymmetric double-gyroid structures has already been demonstrated experimentally in block copolymers \cite{Cowman2015}. Directed self-assembly can induce mechanical strains in the direction of growth \cite{Darling2007} which, according to our symmetry analysis, would automatically lead to the appearance of Weyl points without the need of applying external perturbations. Moreover, gyroid-based systems appear to be unusually resistant against the appearance of cracks when strained \cite{Zhang2017,Werner2013,Robbins2014}, possibly as a result of their three-dimensional co-continuous structure, making them a particularly good fit for our strain-based design.
The size $a$ of the unit cell of the structures obtained by block polymer self-assembly crucially depends on the blocks' molar mass. With current experimental techniques, the accessible unit cell sizes range from a few nanometers to a hundred nanometers \cite{Stefik2015,Meuler2009}. In photonic crystals, this constraint on the unit cell size means that we can expect Weyl points to appear at wavelengths of order $\lambda_{\text{opt}} \sim a / 0.5 \simeq \SI{200}{\nano\meter}$ (or smaller), which are at ultraviolet wavelength. Depending on the materials used in the process, the light frequency may be high enough for the dielectric function not to be constant anymore, but as we have shown, Weyl points can also occur in dispersive photonic crystals. While the direct observation of a Weyl band structure at such frequencies is challenging, such self-assembled photonic crystals could be used in X-ray/XUV optics, for example to realize Veselago lenses, as proposed in reference \cite{Hills2017}.

To generate an optical response in the visible spectrum, it would be interesting to explore hierarchical self-assembly of gyroids using soft building blocks larger than standard polymeric monomers, such as superstructures formed by anisotropic colloids \cite{Glotzer2007,Phillips2012}, or liquid crystalline phases \cite{Aplinc2016}.

\section{Symmetry-driven discovery of self-assembled materials}

Both the possible existence of a three-fold degeneracy and its splitting into Weyl points are predicted by group theory. In the present study, we did not need to make an initial guess of a structure leading to a three-fold degeneracy at $\Gamma$ because we used the well known example of a double gyroid. Note, however, that symmetry considerations can also guide this first step, as they determine in which structures essential degeneracies can exist \cite{BradleyCracknell}, as frequently used both in solid-state physics \cite{Bradlyn2017,Bradlyn2016,Wieder2016} and for classical waves \cite{Lu2014,Saba2017}.
This approach combined with an iterative search through libraries of self-assembled structures could provide an extension of our results to novel systems. We developed open-source Python packages which perform some of the tasks required (see Supporting Informations for details): (i) the script \texttt{space\_group.py} numerically determines the space group of a structure represented as a skeletal graph in presence and absence of mechanical deformations using the open-source library spglib \cite{spglib}, (ii) the script \texttt{irreps.py} numerically determines the irreducible representations of the numerical eigenvectors, (iii) the script \texttt{weyl\_charge.py} computes numerically the charges of the Weyl nodes for an arbitrary band structure using a gauge-invariant method \cite{Luscher1982,Panagiotakopoulos1985,Phillips1990,KingSmith1993,Simon1993,Resta1994,Fukui2005}. 
Figure \ref{summary_flowchart} provides a schematic representation of an automated self-assembled mesostructured material discovery pipeline, which would blend computationally intensive full-wave simulations and the group theoretical tools employed in the present study. In such a scheme as well as in our work, symmetries act as a powerful guide in the wealth of self-assembled structures \cite{Bates1999,Meuler2009,Park2003,Bates2012}, both by identifying candidate systems and determining suitable perturbations to achieve a given response.

\begin{acknowledgments}
M.F. was supported by the NanoFront consortium, a program of the Netherlands Organization for Scientific Research (NWO) that is funded by the Dutch Ministry of Education, Culture and Science (OCW). 
V.V. was primarily supported by the University of Chicago Materials Research Science and Engineering Center, which is funded by the National Science Foundation under award number DMR-1420709. 
K.H. and S.Y.J. were supported by the National Research Foundation of Korea (Grant No. NRF-2016M3D1A1021142).
U.W. would like to thank the National Science foundation for support (DMR-1707836).
\medskip
The code used to compute the photonic band structures, the charges
of the Weyl points, the irreducible representations of the numerical eigenvectors, the
space group of the structures, and the spectra of the effective Hamiltonians is available
on Zenodo at \url{https://doi.org/10.5281/zenodo.1182581}. The Supporting Materials are available as ancillary files at \url{https://arxiv.org/src/1711.11019/anc}.
\end{acknowledgments}

\clearpage

\bibliography{bibliography}

\end{document}